\pdfoutput=0

\documentclass[12pt,a4paper]{article}

\usepackage{ifthen}
\newboolean{pdflatex}
\setboolean{pdflatex}{false}

\newboolean{articletitles}
\setboolean{articletitles}{true}

\newboolean{uprightparticles}
\setboolean{uprightparticles}{false}

\usepackage{lineno}
\usepackage{xspace}

\usepackage{graphicx}
\usepackage{color}
\usepackage{colortbl}
\graphicspath{{./figs/}}

\usepackage{amsmath}
\usepackage{amssymb}
\usepackage{amsfonts}
\usepackage{upgreek}

\newcommand*\patchAmsMathEnvironmentForLineno[1]{%
\expandafter\let\csname old#1\expandafter\endcsname\csname #1\endcsname
\expandafter\let\csname oldend#1\expandafter\endcsname\csname
end#1\endcsname
 \renewenvironment{#1}%
   {\linenomath\csname old#1\endcsname}%
   {\csname oldend#1\endcsname\endlinenomath}%
}
\newcommand*\patchBothAmsMathEnvironmentsForLineno[1]{%
  \patchAmsMathEnvironmentForLineno{#1}%
  \patchAmsMathEnvironmentForLineno{#1*}%
}
\AtBeginDocument{%
\patchBothAmsMathEnvironmentsForLineno{equation}%
\patchBothAmsMathEnvironmentsForLineno{align}%
\patchBothAmsMathEnvironmentsForLineno{flalign}%
\patchBothAmsMathEnvironmentsForLineno{alignat}%
\patchBothAmsMathEnvironmentsForLineno{gather}%
\patchBothAmsMathEnvironmentsForLineno{multline}%
}

\usepackage{hyperref}
\usepackage[all]{hypcap}

\def\lhcb {\mbox{LHCb}\xspace}
\def\ux85 {\mbox{UX85}\xspace}

\ifthenelse{\boolean{uprightparticles}}%
{

 \def\Pmu         {\ensuremath{\upmu}\xspace}                 
 \def\Pnu         {\ensuremath{\upnu}\xspace}                 
                  
 \def\Ppi         {\ensuremath{\uppi}\xspace}

 \def\Ppsi        {\ensuremath{\uppsi}\xspace}

 \def\PDelta      {\ensuremath{\Delta}\xspace}                 
 \def\PXi      {\ensuremath{\Xi}\xspace}                 
 \def\PLambda      {\ensuremath{\Lambda}\xspace}                 
 \def\PSigma      {\ensuremath{\Sigma}\xspace}                 
 \def\POmega      {\ensuremath{\Omega}\xspace}                 
 \def\PUpsilon      {\ensuremath{\Upsilon}\xspace}

 \def\PB      {\ensuremath{\mathrm{B}}\xspace}                 
                  
 \def\PD      {\ensuremath{\mathrm{D}}\xspace}

 \def\PJ      {\ensuremath{\mathrm{J}}\xspace}                 
 \def\PK      {\ensuremath{\mathrm{K}}\xspace}

 \def\Pb      {\ensuremath{\mathrm{b}}\xspace}                 
 \def\Pc      {\ensuremath{\mathrm{c}}\xspace}

 \def\Pi      {\ensuremath{\mathrm{i}}\xspace}

 \def\Pp      {\ensuremath{\mathrm{p}}\xspace}

 \def\Ps      {\ensuremath{\mathrm{s}}\xspace}

}
{

 \def\Pmu         {\ensuremath{\mu}\xspace}                 
 \def\Pnu         {\ensuremath{\nu}\xspace}                 
                  
 \def\Ppi         {\ensuremath{\pi}\xspace}

 \def\Ppsi        {\ensuremath{\psi}\xspace}                 
                  
 \mathchardef\PDelta="7101
 \mathchardef\PXi="7104
 \mathchardef\PLambda="7103
 \mathchardef\PSigma="7106
 \mathchardef\POmega="710A
 \mathchardef\PUpsilon="7107
                  
 \def\PB      {\ensuremath{B}\xspace}                 
                  
 \def\PD      {\ensuremath{D}\xspace}

 \def\PJ      {\ensuremath{J}\xspace}                 
 \def\PK      {\ensuremath{K}\xspace}

 \def\Pb      {\ensuremath{b}\xspace}                 
 \def\Pc      {\ensuremath{c}\xspace}

 \def\Pi      {\ensuremath{i}\xspace}

 \def\Pp      {\ensuremath{p}\xspace}

 \def\Ps      {\ensuremath{s}\xspace}

}

\def\mup        {\ensuremath{\Pmu^+}\xspace}
\def\mun        {\ensuremath{\Pmu^-}\xspace}

\def\neu        {\ensuremath{\Pnu}\xspace}
\def\neub       {\ensuremath{\overline{\Pnu}}\xspace}

\def\squark    {\ensuremath{\Ps}\xspace}

\def\cquark    {\ensuremath{\Pc}\xspace}

\def\bquark    {\ensuremath{\Pb}\xspace}

\def\pion  {\ensuremath{\Ppi}\xspace}

\def\pip   {\ensuremath{\pion^+}\xspace}
\def\pim   {\ensuremath{\pion^-}\xspace}

\def\pipm  {\ensuremath{\pion^\pm}\xspace}
\def\pimp  {\ensuremath{\pion^\mp}\xspace}

\def\kaon  {\ensuremath{\PK}\xspace}
  \def\Kbar  {\kern 0.2em\overline{\kern -0.2em \PK}{}\xspace}

\def\Kz    {\ensuremath{\kaon^0}\xspace}
\def\Kzb   {\ensuremath{\Kbar^0}\xspace}
\def\KzKzb {\ensuremath{\Kz \kern -0.16em \Kzb}\xspace}
\def\Kp    {\ensuremath{\kaon^+}\xspace}
\def\Km    {\ensuremath{\kaon^-}\xspace}
\def\Kpm   {\ensuremath{\kaon^\pm}\xspace}

\def\KpKm  {\ensuremath{\Kp \kern -0.16em \Km}\xspace}

\def\Kstarz  {\ensuremath{\kaon^{*0}}\xspace}
\def\Kstarzb {\ensuremath{\Kbar^{*0}}\xspace}

  \def\Dbar    {\kern 0.2em\overline{\kern -0.2em \PD}{}\xspace}
\def\D       {\ensuremath{\PD}\xspace}

\def\Dz      {\ensuremath{\D^0}\xspace}
\def\Dzb     {\ensuremath{\Dbar^0}\xspace}
\def\DzDzb   {\ensuremath{\Dz {\kern -0.16em \Dzb}}\xspace}
\def\Dp      {\ensuremath{\D^+}\xspace}
\def\Dm      {\ensuremath{\D^-}\xspace}
\def\Dpm     {\ensuremath{\D^\pm}\xspace}
\def\Dmp     {\ensuremath{\D^\mp}\xspace}
\def\DpDm    {\ensuremath{\Dp {\kern -0.16em \Dm}}\xspace}

\def\Dstarzb {\ensuremath{\Dbar^{*0}}\xspace}

\def\Dstarm  {\ensuremath{\D^{*-}}\xspace}
\def\Dstarpm {\ensuremath{\D^{*\pm}}\xspace}

\def\Dspm    {\ensuremath{\D^{\pm}_\squark}\xspace}
\def\Dsmp    {\ensuremath{\D^{\mp}_\squark}\xspace}

\def\B       {\ensuremath{\PB}\xspace}
  \def\Bbar    {\kern 0.18em\overline{\kern -0.18em \PB}{}\xspace}

\def\Bz      {\ensuremath{\B^0}\xspace}
\def\Bzb     {\ensuremath{\Bbar^0}\xspace}
\def\Bu      {\ensuremath{\B^+}\xspace}
\def\Bub     {\ensuremath{\B^-}\xspace}
\def\Bp      {\ensuremath{\Bu}\xspace}
\def\Bm      {\ensuremath{\Bub}\xspace}
\def\Bpm     {\ensuremath{\B^\pm}\xspace}

\def\Bd      {\ensuremath{\B^0}\xspace}
\def\Bs      {\ensuremath{\B^0_\squark}\xspace}
\def\Bds     {\ensuremath{\B^0_{(\squark)}}\xspace}
\def\Bsb     {\ensuremath{\Bbar^0_\squark}\xspace}

\def\jpsi     {\ensuremath{{\PJ\mskip -3mu/\mskip -2mu\Ppsi\mskip 2mu}}\xspace}

  \def\Y#1S{\ensuremath{\PUpsilon{(#1S)}}\xspace}

\def\FourS {\Y4S}

\def\proton      {\ensuremath{\Pp}\xspace}
\def\antiproton  {\ensuremath{\overline \proton}\xspace}

\def\Lbar {\ensuremath{\kern 0.1em\overline{\kern -0.1em\PLambda}}\xspace}

\def\Lbbar   {\ensuremath{\Lbar^0_\bquark}\xspace}

\def\to                 {\ensuremath{\rightarrow}\xspace}

\def\CP                {\ensuremath{C\!P}\xspace}

\def\AT#1     {\ensuremath{A_{\mathrm{T}}^{#1}}\xspace}

\def\C#1      {\ensuremath{\mathcal{C}_{#1}}\xspace}                       
\def\Cp#1     {\ensuremath{\mathcal{C}_{#1}^{'}}\xspace}                    
\def\Ceff#1   {\ensuremath{\mathcal{C}_{#1}^{\mathrm{(eff)}}}\xspace}        
\def\Cpeff#1  {\ensuremath{\mathcal{C}_{#1}^{'\mathrm{(eff)}}}\xspace}       
\def\Ope#1    {\ensuremath{\mathcal{O}_{#1}}\xspace}                       
\def\Opep#1   {\ensuremath{\mathcal{O}_{#1}^{'}}\xspace}

\newcommand{\tev}{\ensuremath{\mathrm{\,Te\kern -0.1em V}}\xspace}
\newcommand{\gev}{\ensuremath{\mathrm{\,Ge\kern -0.1em V}}\xspace}
\newcommand{\mev}{\ensuremath{\mathrm{\,Me\kern -0.1em V}}\xspace}
\newcommand{\kev}{\ensuremath{\mathrm{\,ke\kern -0.1em V}}\xspace}
\newcommand{\ev}{\ensuremath{\mathrm{\,e\kern -0.1em V}}\xspace}
\newcommand{\gevc}{\ensuremath{{\mathrm{\,Ge\kern -0.1em V\!/}c}}\xspace}
\newcommand{\mevc}{\ensuremath{{\mathrm{\,Me\kern -0.1em V\!/}c}}\xspace}
\newcommand{\gevcc}{\ensuremath{{\mathrm{\,Ge\kern -0.1em V\!/}c^2}}\xspace}
\newcommand{\gevgevcccc}{\ensuremath{{\mathrm{\,Ge\kern -0.1em V^2\!/}c^4}}\xspace}
\newcommand{\mevcc}{\ensuremath{{\mathrm{\,Me\kern -0.1em V\!/}c^2}}\xspace}

\def\mm   {\ensuremath{\rm \,mm}\xspace}

\def\mum  {\ensuremath{\,\upmu\rm m}\xspace}

\def\invfb   {\ensuremath{\mbox{\,fb}^{-1}}\xspace}

\newcommand{\chisq}{\ensuremath{\chi^2}\xspace}

\def\gsim{{~\raise.15em\hbox{$>$}\kern-.85em
          \lower.35em\hbox{$\sim$}~}\xspace}
\def\lsim{{~\raise.15em\hbox{$<$}\kern-.85em
          \lower.35em\hbox{$\sim$}~}\xspace}

\def\sPlot{\mbox{\em sPlot}}

\def\pt         {\mbox{$p_{\rm T}$}\xspace}

\def\evtgen     {\mbox{\textsc{EvtGen}}\xspace}
\def\pythia     {\mbox{\textsc{Pythia}}\xspace}

\def\geant      {\mbox{\textsc{Geant4}}\xspace}

\def\gauss      {\mbox{\textsc{Gauss}}\xspace}

\def\photos     {\mbox{\textsc{Photos}}\xspace}

\def\tell1  {TELL1\xspace}
\def\ukl1   {UKL1\xspace}

\usepackage{cite} 
\usepackage{mciteplus}

\begin{document}

\renewcommand{\thefootnote}{\fnsymbol{footnote}}
\setcounter{footnote}{1}

\begin{titlepage}
\pagenumbering{roman}

\vspace*{-1.5cm}
\centerline{\large EUROPEAN ORGANIZATION FOR NUCLEAR RESEARCH (CERN)}
\vspace*{1.5cm}
\hspace*{-0.5cm}
\begin{tabular*}{\linewidth}{lc@{\extracolsep{\fill}}r}
\ifthenelse{\boolean{pdflatex}}
{\vspace*{-2.7cm}\mbox{\!\!\!\includegraphics[width=.14\textwidth]{figs/lhcb-logo.pdf}} & &}%
{\vspace*{-1.2cm}\mbox{\!\!\!\includegraphics[width=.12\textwidth]{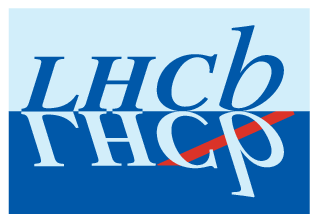}} & &}%
\\
 & & CERN-PH-EP-2013-070 \\  
 & & LHCb-PAPER-2013-022 \\  
 & & June 4, 2013 \\
 & & \\
\end{tabular*}

\vspace*{0.5cm}

{\bf\boldmath\huge
\begin{center}
  Measurement of the branching fractions of the decays $\Bs \to \Dzb\Km\pip$ and  $\Bd \to \Dzb\Kp\pim$
\end{center}
}

\vspace*{0.5cm}

\begin{center}
The LHCb collaboration\footnote{Authors are listed on the following pages.}
\end{center}\

\begin{abstract}
  \noindent
  The first observation of the decay $\Bs \to \Dzb K^-\pi^+$ is reported.
  The analysis is based on a data sample, corresponding to an integrated luminosity of $1.0 \invfb$ of $pp$ collisions, collected with the LHCb detector.
  The branching fraction relative to that of the topologically similar decay $\Bd \to \Dzb \pi^+\pi^-$ is measured to be
  $$
  \frac{
    {\cal B}\left(\Bs\to \Dzb K^-\pi^+\right)}{
    {\cal B}\left(\Bd \to \Dzb \pi^+\pi^-\right)} = 1.18 \pm 0.05\,\text{(stat.)} \pm 0.12\,\text{(syst.)} \, .
  $$
  In addition, the relative branching fraction of the decay $\Bd \to \Dzb K^+\pi^-$ is measured to be
  $$
  \frac{
  {\cal B}\left(\Bd \to \Dzb K^+\pi^-\right)}{
  {\cal B}\left(\Bd \to \Dzb \pi^+\pi^-\right)} = 0.106 \pm 0.007\,\text{(stat.)} \pm 0.008\,\text{(syst.)} \, .
  $$
\end{abstract}

\vspace*{0.5cm}

\begin{center}
  Submitted to Phys.\ Rev.\ D.
\end{center}

\vspace{\fill}

{\footnotesize 
\centerline{\copyright~CERN on behalf of the \lhcb collaboration, license \href{http://creativecommons.org/licenses/by/3.0/}{CC-BY-3.0}.}}
\vspace*{2mm}

\end{titlepage}

\newpage
\setcounter{page}{2}
\mbox{~}
\newpage

\centerline{\large\bf LHCb collaboration}
\begin{flushleft}
\small
R.~Aaij$^{40}$, 
C.~Abellan~Beteta$^{35,n}$, 
B.~Adeva$^{36}$, 
M.~Adinolfi$^{45}$, 
C.~Adrover$^{6}$, 
A.~Affolder$^{51}$, 
Z.~Ajaltouni$^{5}$, 
J.~Albrecht$^{9}$, 
F.~Alessio$^{37}$, 
M.~Alexander$^{50}$
% ...

{\footnotesize \it
% ...
$ ^{5}$Clermont Universit\'{e}, Universit\'{e} Blaise Pascal, CNRS/IN2P3, LPC, Clermont-Ferrand, France\\
$ ^{6}$CPPM, Aix-Marseille Universit\'{e}, CNRS/IN2P3, Marseille, France\\
% ...
$ ^{9}$Fakult\"{a}t Physik, Technische Universit\"{a}t Dortmund, Dortmund, Germany\\
% ...
$ ^{35}$Universitat de Barcelona, Barcelona, Spain\\
$ ^{36}$Universidad de Santiago de Compostela, Santiago de Compostela, Spain\\
$ ^{37}$European Organization for Nuclear Research (CERN), Geneva, Switzerland\\
% ...
$ ^{40}$Nikhef National Institute for Subatomic Physics, Amsterdam, The Netherlands\\
% ...
$ ^{45}$H.H. Wills Physics Laboratory, University of Bristol, Bristol, United Kingdom\\
% ...
$ ^{50}$School of Physics and Astronomy, University of Glasgow, Glasgow, United Kingdom\\
$ ^{51}$Oliver Lodge Laboratory, University of Liverpool, Liverpool, United Kingdom\\
% ...
$ ^{n}$LIFAELS, La Salle, Universitat Ramon Llull, Barcelona, Spain\\
% ...
}
\end{flushleft}

\cleardoublepage

\renewcommand{\thefootnote}{\arabic{footnote}}
\setcounter{footnote}{0}

\pagestyle{plain}
\setcounter{page}{1}
\pagenumbering{arabic}

\section{Introduction}

The precise measurement of the angle $\gamma$ of the CKM Unitarity Triangle~\cite{Cabibbo:1963yz,Kobayashi:1973fv} is one of the primary objectives in contemporary flavour physics.
Measurements from the experiments BaBar, Belle and LHCb are based mainly on studies of \mbox{$\Bp \to D\Kp$} decays, where the notation $\D$ implies that the neutral $\D$ meson is an admixture of $\Dz$ and $\Dzb$ states.
Each experiment currently gives constraints on $\gamma$ with a precision of $\sim 15^\circ$~\cite{:2013zd,Trabelsi:2013uj,LHCb-PAPER-2013-020}.
Significant reduction of this uncertainty is well motivated and the use of additional channels to further improve the precision is of great interest. 

The decay \mbox{$\Bd \to \D\Kp\pim$}, including the resonant contribution from \mbox{$\Bd \to \D\Kstarz$}, is one of the modes with the potential to make significant impact on the overall determination of $\gamma$~\cite{Gronau:2002mu}.
A first measurement of \CP observables in \mbox{$\Bd \to \D\Kstarz$} decays has been reported by LHCb~\cite{LHCb-PAPER-2012-042}.
This decay is particularly sensitive to $\gamma$ owing to the interference of \mbox{$b\to c\bar{u}s$} and \mbox{$b\to u\bar{c}s$} amplitudes, 
which for this decay are of similar magnitude.
It has been noted that an amplitude analysis of \mbox{$\Bd \to \D\Kp\pim$} decays can further improve the sensitivity and also resolve the ambiguities in the result~\cite{Gershon:2008pe,Gershon:2009qc}. 

\begin{figure}[!b]
  \centering
  \includegraphics[width=0.330\textwidth]{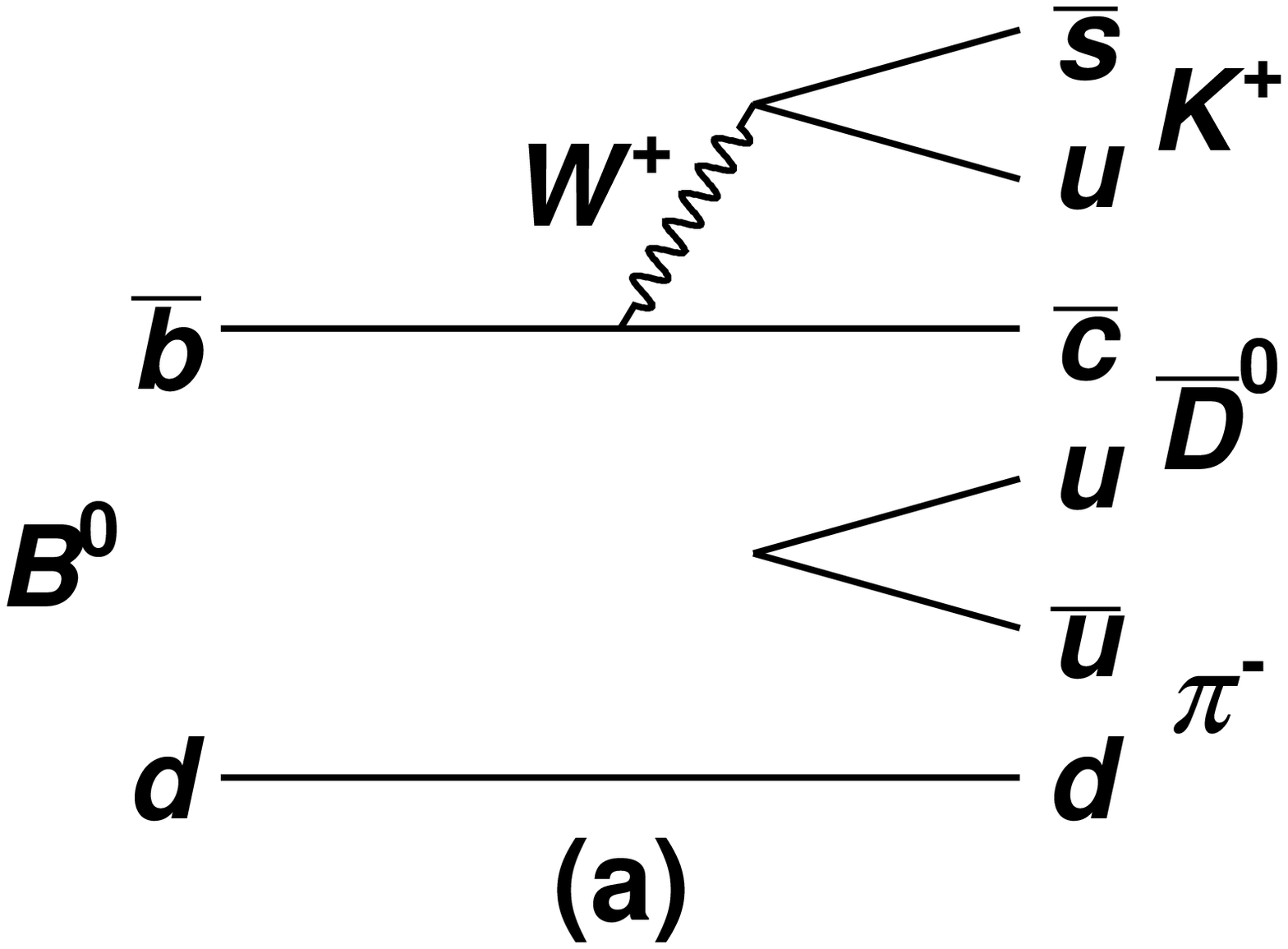}
  \includegraphics[width=0.330\textwidth]{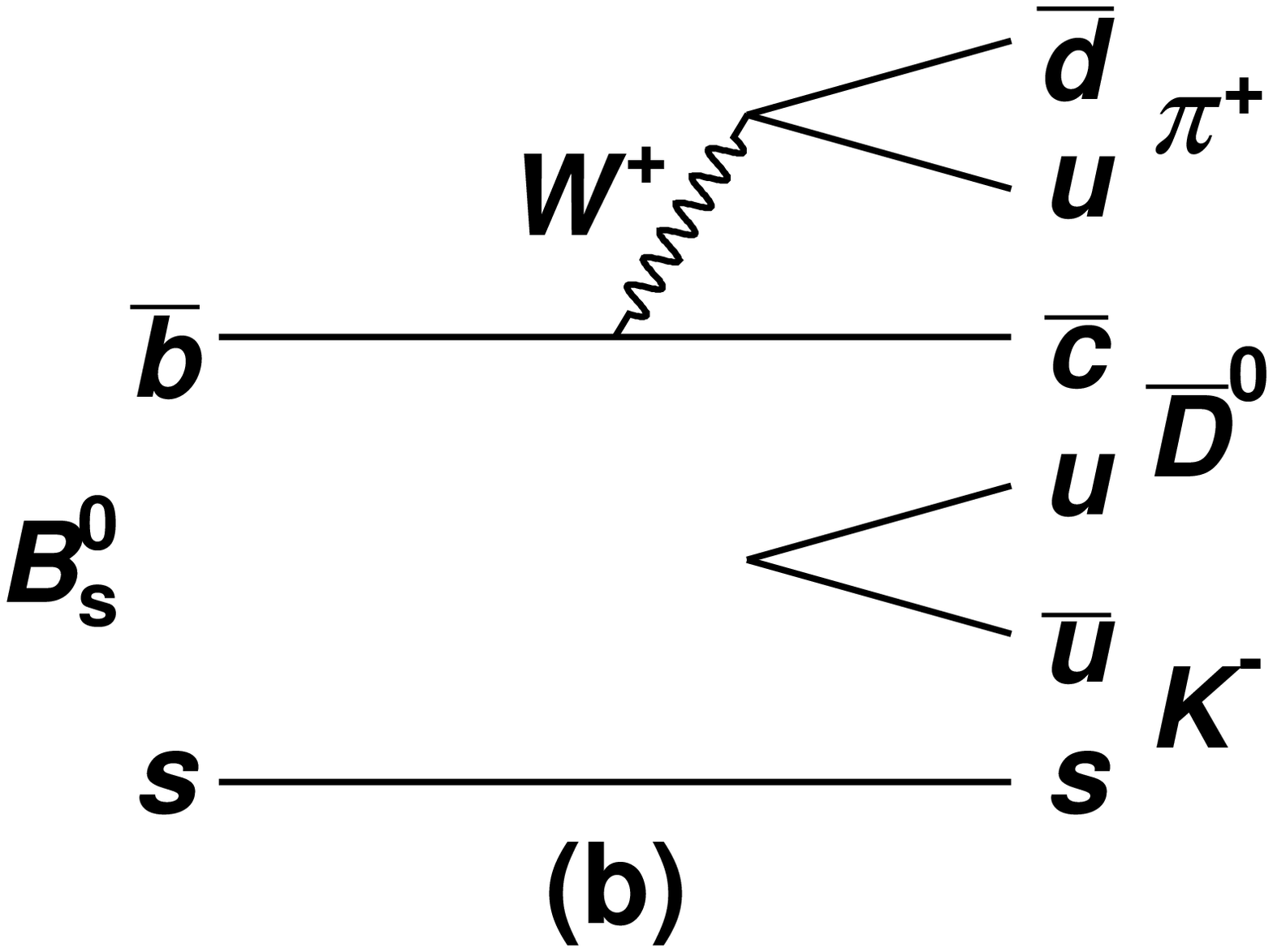}
  \caption{
    \small{Decay diagrams for (a) favoured \mbox{$\Bd \to \Dzb\Kp\pim$} decays and (b) favoured \mbox{$\Bs \to \Dzb\Km\pip$} decays.}
  }
  \label{fig:feynman}
\end{figure}

The decays \mbox{$\Bd \to \Dzb\Kp\pim$} and \mbox{$\Bs \to \Dzb\Km\pip$} can be mediated by the decay diagrams shown in Fig.~\ref{fig:feynman}.
Both \Bd and \Bs decays are flavour-specific, with the charge of the kaon identifying the flavour of the decaying \B meson, though the charges are opposite in the two cases. 
In addition to these colour-allowed tree-level diagrams, colour-suppressed tree-level diagrams contribute to \mbox{$\Bds\to \Dzb K\pi$} decays ($K\pi$ denotes the sum over both charge combinations).
Both colour-allowed and colour-suppressed diagrams contribute to the CKM-suppressed \mbox{$\Bds\to \Dz K\pi$} modes.

A first study of the decay \mbox{$\Bd \to \Dzb\Kp\pim$} has been performed by BaBar~\cite{Aubert:2005yt}, giving a branching fraction measurement
${\cal B}\left( \Bd \to \Dzb\Kp\pim \right) = (88 \pm 15 \pm 9) \times 10^{-6}$,
where the contribution from the \mbox{$\Bd \to \Dstarm\Kp$} decay is excluded.
There is no previous branching fraction measurement for the inclusive three-body process \mbox{$\Bs \to \Dzb\Km\pip$}, although that of the resonant contribution $\Dzb\Kstarzb$ has been measured by LHCb~\cite{LHCb-PAPER-2011-008}.
Since the \mbox{$\Bs \to \Dzb\Km\pip$} and the related \mbox{$\Bs \to \Dstarzb\Km\pip$} decays form potentially serious backgrounds to the $\mbox{\Bd \to \D\Kp\pim}$ channel, measurements of their properties will be necessary to reduce systematic uncertainties in the determination of $\gamma$.

In this paper the results of a study of neutral \B meson decays to $\Dzb K\pi$, including inspections of their Dalitz plot distributions, are presented.
The $\Dzb\Kp\pim$ and $\Dzb\Km\pip$ final states are combined, and the inclusion of charge conjugate processes is implied throughout the paper.  
In order to reduce systematic uncertainties in the measurements, the topologically similar decay $\Dzb \pip\pim$, which has been studied in detail previously~\cite{Kuzmin:2006mw,:2010ip}, is used as a normalisation channel.
In this paper, $D\pi\pi$ denotes the $\Dzb\pip\pim$ final state and $DK\pi$ denotes the sum over the $\Dzb\Kp\pim$ and $\Dzb\Km\pip$ final states.
The neutral $\D$ meson is reconstructed using the \mbox{$\Dzb \to \Kp\pim$} final state; therefore the signal yields measured include small contributions from \mbox{$\Dz \to \Kp\pim$} decays, but such contributions are expected to be small and are neglected hereafter.
The analysis uses a data sample, corresponding to an integrated luminosity of $1.0 \invfb$ of $pp$ collisions at a centre-of-mass energy of $7 \tev$, collected with the LHCb detector during 2011.

\section{Detector, trigger and selection}

The \lhcb detector~\cite{Alves:2008zz} is a single-arm forward
spectrometer covering the \mbox{pseudorapidity} range $2<\eta <5$,
designed for the study of particles containing \bquark or \cquark
quarks. The detector includes a high precision tracking system
consisting of a silicon-strip vertex detector surrounding the $pp$
interaction region, a large-area silicon-strip detector located
upstream of a dipole magnet with a bending power of about
$4{\rm\,Tm}$, and three stations of silicon-strip detectors and straw
drift tubes placed downstream. The combined tracking system provides
a momentum measurement with relative uncertainty that varies from 0.4\,\% at 5\gevc to
0.6\,\% at 100\gevc, and impact parameter (IP) resolution of 20\mum for
tracks with high transverse momentum ($\pt$). Charged hadrons are identified
using two ring-imaging Cherenkov (RICH) detectors~\cite{LHCb-DP-2012-003}. Photon, electron and
hadron candidates are identified by a calorimeter system consisting of
scintillating-pad and preshower detectors, an electromagnetic
calorimeter and a hadronic calorimeter. Muons are identified by a
system composed of alternating layers of iron and multiwire
proportional chambers. 

The LHCb trigger~\cite{LHCb-DP-2012-004} consists of a
hardware stage, based on information from the calorimeter and muon
systems, followed by a software stage that applies a full event
reconstruction.
In this analysis, signal candidates are accepted if one of the final state particles created a cluster in the hadronic calorimeter with sufficient transverse energy to fire the hardware trigger. 
Events that are triggered at the hardware level by another particle in the event are also retained.

The software trigger requires a two-, three- or four-track
secondary vertex with a high sum of the transverse momentum, \pt, of
the tracks and a significant displacement from the primary $pp$
interaction vertices~(PVs). At least one track should have $\pt >
1.7\gevc$ and impact parameter \chisq, $\chisq_{\rm IP}$, with respect to the
primary interaction greater than 16.
The $\chisq_{\rm IP}$ is the difference between the \chisq of the PV reconstruction with and without the considered track.
A multivariate algorithm~\cite{Gligorov:2012qt} is used for the identification of secondary vertices consistent with the decay of a \bquark hadron.

Candidates that satisfy the software trigger selection and are consistent with the decay chain \mbox{$\Bds \to \Dzb \Kpm\pimp$}, \mbox{$\Dzb \to \Kp\pim$} are selected, with requirements similar to those in the LHCb study of the decay \mbox{$\Bds \to \Dzb\Kp\Km$}~\cite{LHCb-PAPER-2012-018}.
The $\Dzb$ candidate invariant mass is required to satisfy \mbox{$1844 < m_{K\pi} < 1884 \mevcc$}.
Tracks are required to be consistent with either the kaon or pion hypothesis, as appropriate, based on particle identification (PID) information primarily from the RICH detectors~\cite{LHCb-DP-2012-003}.
All other selection criteria were tuned on the $\Dzb \pip\pim$ channel.
The large yield available for the \mbox{$\Bz \to \Dzb \pip\pim$} normalisation sample allows the selection to be based on data, though the efficiencies are determined using simulated events.
In the simulation, $pp$ collisions are generated using
\pythia~6.4~\cite{Sjostrand:2006za} with a specific \lhcb
configuration~\cite{LHCb-PROC-2010-056}.  Decays of hadronic particles
are described by \evtgen~\cite{Lange:2001uf} in which final state
radiation is generated using \photos~\cite{Golonka:2005pn}. The 
interaction of the generated particles with the detector and its
response are implemented using the \geant
toolkit~\cite{Allison:2006ve, *Agostinelli:2002hh} as described in
Ref.~\cite{LHCb-PROC-2011-006}.

Loose selection requirements are applied to obtain a visible signal peak in the $\Dzb \pip\pim$ normalisation channel.
The selection includes criteria on the quality of the tracks forming the signal candidate, their $p$, $\pt$ and $\chisq_{\rm IP}$. 
Requirements are also placed on the corresponding variables for candidate composite particles ($\Dzb$, $\Bds$) together with restrictions on the consistency of the decay fit ($\chisq_{\rm vertex}$), the flight distance significance ($\chisq_{\rm flight}$),
and the cosine of the angle between the momentum vector and the line joining the PV under consideration to the $\Bds$ vertex ($\cos \theta_{\rm dir}$)~\cite{LHCb-PAPER-2011-008}.

A boosted decision tree (BDT)~\cite{Breiman} that identifies \mbox{$\Dzb\to \Kp\pim$} candidates is used to suppress backgrounds from \bquark-hadron decays to final states that do not contain charmed particles and backgrounds where the $\Dzb$ does not decay to the $\Kp\pim$ final state. 
This ``$\Dz$ BDT''~\cite{LHCb-PAPER-2012-025,LHCb-PAPER-2012-050} is trained using a large high-purity sample obtained from \mbox{$\Bp \to \Dzb\pip$} decays. 
The BDT takes advantage of the kinematic similarity of all \bquark-hadron decays and avoids using any topological information from the $\Bds$ decay. 
Properties of the $\Dzb$ candidate and its daughter tracks, containing kinematic, track quality, vertex and PID information, are used to train the BDT. 

Further discrimination between signal and background categories is achieved by calculating weights, using the \sPlot\ technique~\cite{Pivk:2004ty}, 
for the remaining $\Dzb\pip\pim$ candidates. 
The weights are based on a simplified fit to the $B$ candidate invariant mass distribution from the $\D\pi\pi$ data sample.
The weights are used to train a neural network~\cite{Feindt2006190} to maximise the separation between the categories.
A total of 10 variables are used in the network. They include the $\pt$, $\chisq_{\rm IP}$, $\chisq_{\rm vertex}$, $\chisq_{\rm flight}$ and $\cos \theta_{\rm dir}$ of the $\Bds$ candidate, the output of the $\Dz$ BDT and the $\chisq_{\rm IP}$ of the two pion tracks that originate from the $\Bds$ vertex.
The $\pt$ asymmetry and track multiplicity in a cone with half-angle of 1.5 units in the plane of pseudorapidity and azimuthal angle (measured in radians)~\cite{LHCB-PAPER-2012-001} around the $\Bds$ candidate flight direction are also used.
The input quantities to the neural network only depend weakly on the kinematics of the $\Bds$ decay.
A requirement on the network output is imposed that reduces the combinatorial background by an order of magnitude while retaining about 70\,\% of the signal.

To improve the $\Bds$ candidate invariant mass resolution, the four-momenta of the tracks from the $\Dzb$ candidate are adjusted~\cite{Hulsbergen:2005pu} so that their combined invariant mass matches the world average value~\cite{PDG2012}.  
An additional $\Bds$ mass constraint is applied in the calculation of the Dalitz plot coordinates, $m^{2}(DK)$ and $m^2(D\pi)$, which are used in the determination of event-by-event efficiencies.  
The coordinates are calculated twice: once each with a $\Bd$ and a $\Bs$ mass constraint.
A small fraction ($\sim 1\,\%$ within the fitted mass range) of candidates with invariant masses far from the $\Bds$ peak fail one or both of these mass-constrained fits, and are removed from the analysis.

To remove the large background from \mbox{$\Bd \to \Dstarm\pip$} decays, candidates in both samples are rejected if the mass difference $m_{D\pi}$--$m_{D}$ (for either pion charge in the combinations $\Dzb\pip\pim$ and $\Dzb K\pi$) lies within $\pm 2.5 \mevcc$ of the nominal $\Dstarm$--$\Dzb$ mass difference~\cite{PDG2012}.
Candidates in the $DK\pi$ sample are also rejected if the mass difference $m_{DK}$--$m_{D}$ calculated under the pion mass hypothesis satisfies the same criterion.
A potential background contribution from \mbox{$\Bs\to\Dmp\Kpm$} decays is removed by requiring that the pion from the $\Dzb$ candidate together with the kaon and the pion do not form an invariant mass in the range $1850$--$1885 \mevcc$. 
Further $DK\pi$ candidates are rejected by requiring that the kaon from the $\Dzb$ candidate together with the kaon and the pion do not form an invariant mass in the range $1955$--$1975 \mevcc$, which removes potential background from \mbox{$\Bs\to\Dsmp\pipm$} decays. 
A muon veto is applied to all four final state tracks to remove potential background from \mbox{$\Bds\to\jpsi\Kstarz$} decays and $\Dzb$ candidates are required to travel at least $1\mm$ from the $\Bds$ decay vertex to remove charmless backgrounds that survive the $\Dz$ BDT requirement. 

Candidates are retained for further analysis if they have an invariant mass in the range $5150$--$5600 \mevcc$ for $\D\pi\pi$ or $5200$--$5600 \mevcc$ for $\D K\pi$.
After all selection requirements are applied, fewer than 1\,\% of events with at least one candidate also contain a second candidate.
Such multiple candidates are retained and treated in the same manner as other candidates; the associated systematic uncertainty is negligible.

\section{Determination of signal yields}

The signal yields are obtained from unbinned maximum likelihood fits to the invariant mass distributions.
In addition to signal contributions and combinatorial background, candidates may be formed from misidentified or partially reconstructed \bquark-hadron decays.
Contributions from partially reconstructed decays are reduced by the lower bounds on the invariant mass regions used in the fits.
Sources of misidentified backgrounds are investigated using simulation.  
Most potential sources are found to have broad invariant mass distributions, and are absorbed in the combinatorial background shapes used in the fits described below.  
Backgrounds from \mbox{$\Lbbar \to \Dzb \antiproton \pip$}~\cite{LHCb-CONF-2011-036} and \mbox{$\Bd \to \Dzb \pip\pim$} decays may, however, give contributions with distinctive shapes in the mass distributions of $\D\pi\pi$ and $\D K\pi$ candidates, respectively, and are therefore explicitly modelled in the fits.

\begin{figure}[!tb]
  \centering
  \includegraphics[width=0.495\textwidth]{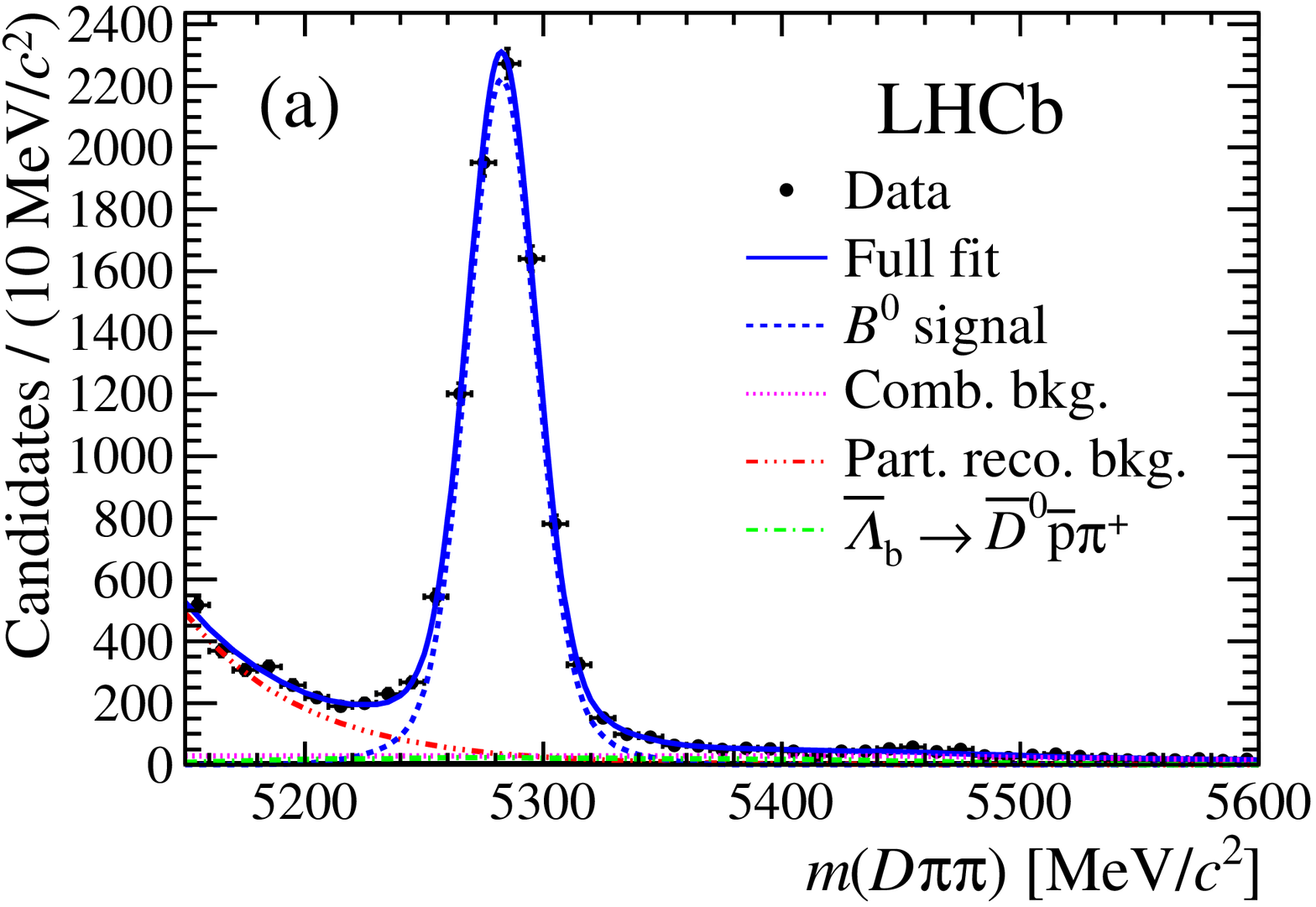}
  \includegraphics[width=0.495\textwidth]{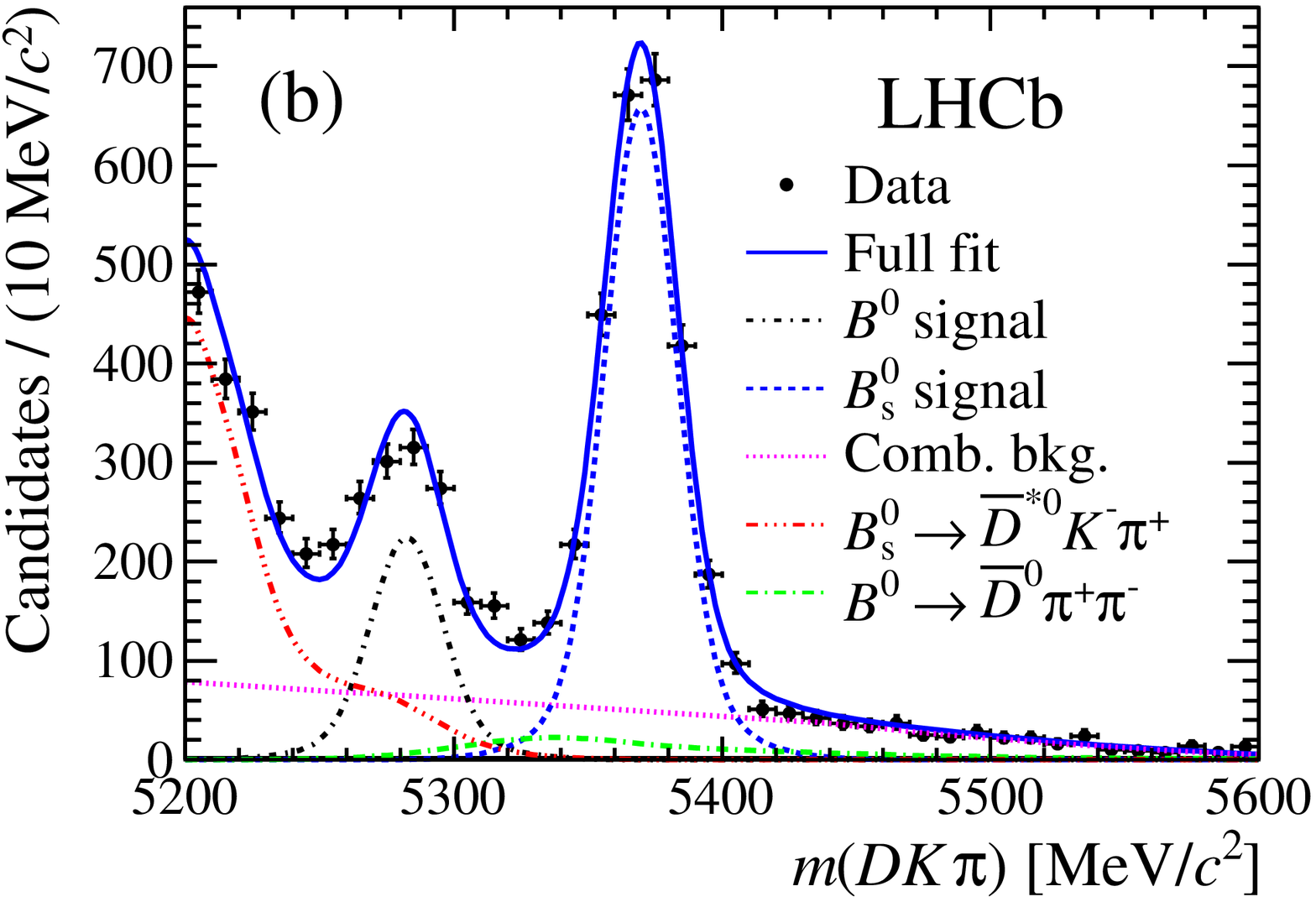}
  \caption{
    \small{Fits to the $\Bds$ candidate invariant mass distributions for the (a)~$D\pi\pi$ and (b)~$DK\pi$ samples. 
    Data points are shown in black, the full fitted PDFs as solid blue lines and the components as detailed in the legends.}
  }
  \label{fig:fits}
\end{figure}

The $D\pi\pi$ fit includes a double Gaussian shape to describe the signal, where the two Gaussian functions share a common mean, together with an exponential component for partially reconstructed background, and a probability density function (PDF) for \mbox{$\Lbbar \to \Dzb \antiproton \pip$} decays.
This PDF is modelled using a smoothed non-parametric function obtained from simulated data, reweighted so that the $\Dzb\pip$ invariant mass distribution matches that observed in data.
The shape of the combinatorial background is essentially linear, but is multiplied by a function that accounts for the fact that candidates with high invariant masses are more likely to fail the $\Bds$ mass constrained fit.
There are ten free parameters in the $\D\pi\pi$ fit:
the double Gaussian peak position, the widths of the two Gaussian shapes and the relative normalisation of the two Gaussian functions,
the linear slope of the combinatorial background, the exponential shape parameter of the partially reconstructed background, 
and the yields of the four categories.
The result of the fit to the $D\pi\pi$ candidates is shown in Fig.~\ref{fig:fits}(a) and yields $8558 \pm 134$ \mbox{$\Bd \to \Dzb\pip\pim$} decays.

The $DK\pi$ fit includes a second double Gaussian component to account for the presence of both $\Bd$ and $\Bs$ decays. The peaking background PDF for \mbox{$\Bd \to \Dzb\pip\pim$} decays is modelled using a smoothed non-parametric function derived from simulation, reweighted in the same way as described for \mbox{$\Lbbar \to \Dzb \antiproton \pip$} decays above. 
The dominant partially reconstructed backgrounds in the $\D K\pi$ fit are from $\Bs$ decays and these extend into the $\Bd$ signal region. 
Instead of an exponential component, a background PDF for \mbox{$\Bs \to \Dstarzb \Km\pip$} decays is included, modelled using a smoothed non-parametric function obtained from simulation. 
Studies using simulated data show that this function can account for all resonant contributions to the \mbox{$\Bs \to \Dstarzb \Km\pip$} final state.
The function describing the combinatorial background has the same form as for the $\D\pi\pi$ fit.
The $DK\pi$ fit has eight free parameters;
the parameters of the double Gaussian functions are constrained to be identical for the $\Bd$ and $\Bs$ signals, with an offset in their mean values fixed to the known $\Bd$--$\Bs$ mass difference~\cite{PDG2012}.
The relative width of the broader to the narrower Gaussian component and the relative normalisation of the two Gaussian functions are constrained within their uncertainties to the values obtained in simulation. 
The result of the fit is shown in Fig.~\ref{fig:fits}(b) and yields $815 \pm 55$ \mbox{$\Bd \to \Dzb\Kp\pim$} and $2391 \pm 81$ \mbox{$\Bs \to \Dzb\Km\pip$} decays.
All background yields in both fits are consistent with their expectations within uncertainties, based on measured or predicted production rates and branching fractions and background rejection factors determined from simulations.

\section{Calculation of branching fraction ratios}

The ratios of branching fractions are obtained after applying event-by-event efficiencies as a function of the Dalitz plot position.
The branching fraction for the $\Bd \to \Dzb\Kp\pim$ decay is determined as
\begin{equation}
  \label{eqn:bfratio}
  R_{\Bd} \equiv
  \frac{
    {\cal B}\left(\Bd \to \Dzb\Kp\pim\right)}{
    {\cal B}\left(\Bd \to \Dzb\pip\pim\right)} = 
  \frac{
    N^{\rm corr}(\Bd \to \Dzb\Kp\pim)}{
    N^{\rm corr}(\Bd \to \Dzb\pip\pim)}\, ,
\end{equation}
and the branching fraction of the $\Bs \to \Dzb\Km\pip$ mode is determined as
\begin{equation}
  R_{\Bs} \equiv
  \frac{
    {\cal B}\left(\Bs \to \Dzb\Km\pip\right)}{
    {\cal B}\left(\Bd \to \Dzb\pip\pim\right)} = 
  \left( \frac{f_s}{f_d} \right)^{-1}
  \frac{N^{\rm corr}(\Bs \to \Dzb\Km\pip)}{N^{\rm corr}(\Bd \to \Dzb\pip\pim)} \, ,
\end{equation}
where the efficiency corrected yield is $N^{\rm corr} = \sum_i W_i / \epsilon^{\rm tot}_i$.
Here the index $i$ runs over all candidates in the fit range,
$W_i$ is the signal weight for candidate $i$, determined using the procedure described in Ref.~\cite{Pivk:2004ty}, from the fits shown in Fig.~\ref{fig:fits}
and $\epsilon^{\rm tot}_i$ is the efficiency for candidate $i$ as a function of its Dalitz plot position.
The ratio of fragmentation fractions is 
$f_s/f_d = 0.256 \pm 0.020$~\cite{LHCb-PAPER-2012-037}.
The statistical uncertainty on the branching fraction ratio incorporates the effects of the shape parameters that are allowed to vary in the fit and the dilution due to event weighting.
Most potential systematic effects cancel in the ratio. 

The PID efficiency is measured using 
a control sample of \mbox{$\Dstarm \to \Dzb \pim,\,\Dzb \to \Kp\pim$} decays to obtain background-subtracted efficiency tables for kaons and pions as a function of their $p$ and $\pt$~\cite{LHCb-PAPER-2012-002,LHCb-DP-2012-003}.
The kinematic properties of the particles in signal decays are obtained from simulation in which events are uniformly distributed across the phase space, allowing the PID efficiency for each event to be obtained from the tables, while taking into account the correlation between the $p$ and $\pt$ values of the two tracks.
The other contributions to the efficiency (detector acceptance, selection criteria and trigger effects) are determined from phase space simulation, and validated using data.
All are found to be approximately constant across the Dalitz plane, apart from some modulations seen near the kinematic boundaries and, for the $DK\pi$ channels, a variation caused by different PID requirements on the pion and the kaon.
The efficiency for each mode, averaged across the Dalitz plot, is given in Table~\ref{tab:efficiencies} together with the contributions from geometrical acceptance, trigger and selection requirements and particle identification.

\begin{table}[!tb]
  \caption{\small
    Summary of the efficiencies for $D\pi\pi$ and $DK\pi$ in phase space simulation. 
    Contributions from geometrical acceptance ($\epsilon^{\rm geom}$), trigger and selection requirements ($\epsilon^{\rm trig\&sel}$) and particle identification ($\epsilon^{\rm PID}$) are shown.
    The geometrical acceptance is evaluated for \B mesons produced within the detector acceptance.
    Values given are in percent.
  }
  \label{tab:efficiencies}
  \centering
  \begin{tabular}{lccc}
    & $\Bd\to D\pi\pi$ & $\Bd\to DK\pi$ & $\Bs\to DK\pi$ \\
    \hline
    $\epsilon^{\rm geom}$ & 44.7\phantom{0} & 46.6\phantom{0} & 46.5\phantom{0} \\
    $\epsilon^{\rm trig\&sel}$ & \phantom{0}1.32 & \phantom{0}1.25 & \phantom{0}1.25 \\
    $\epsilon^{\rm PID}$ & 89.3\phantom{0} & 74.8\phantom{0} & 75.0\phantom{0} \\
    \hline
    $\epsilon^{\rm tot}$ & \phantom{0}0.53 & \phantom{0}0.44 & \phantom{0}0.44 \\
  \end{tabular}
\end{table}

\begin{figure}[tb]
  \centering
  \includegraphics[width=0.325\textwidth]{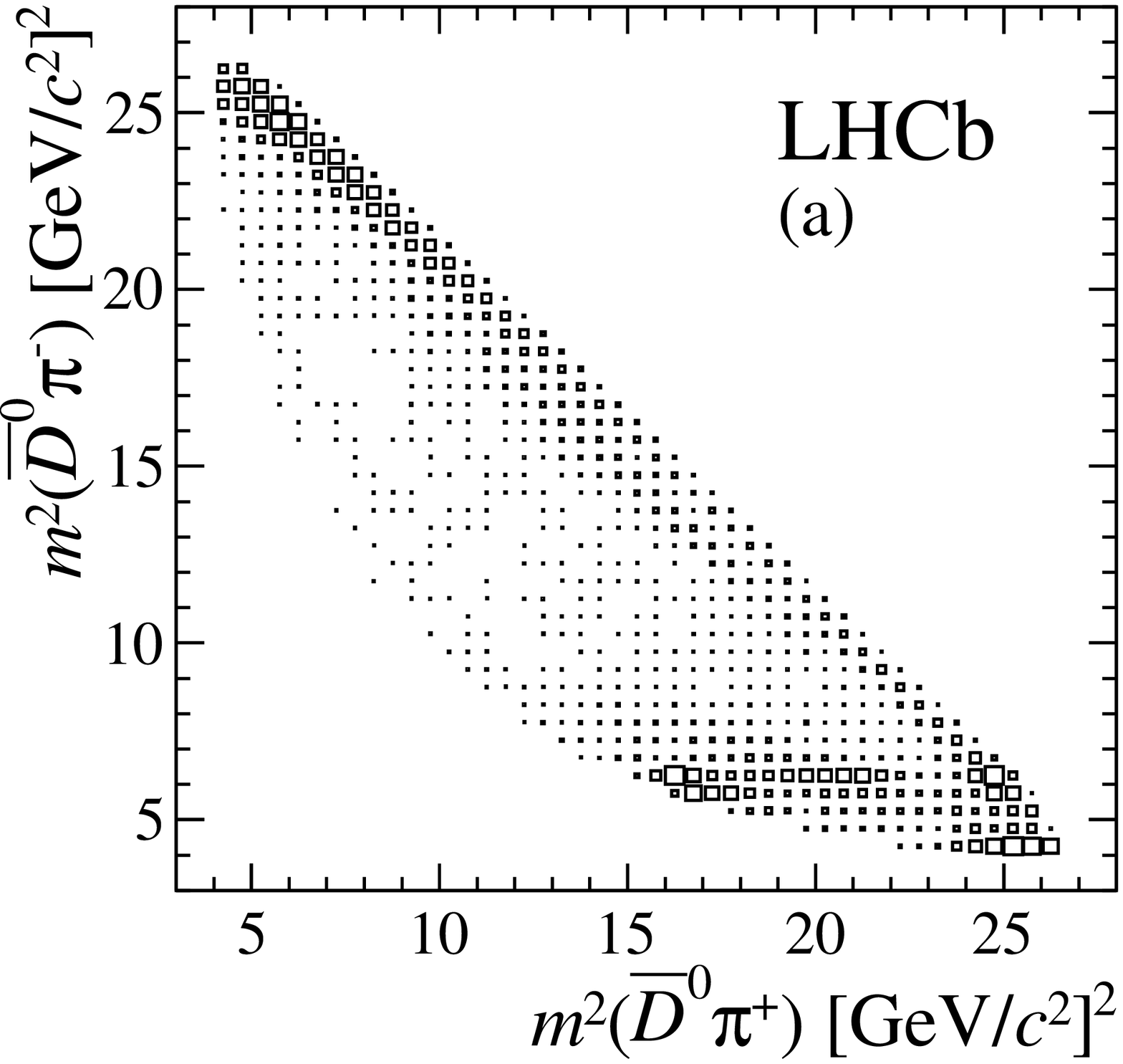}
  \includegraphics[width=0.325\textwidth]{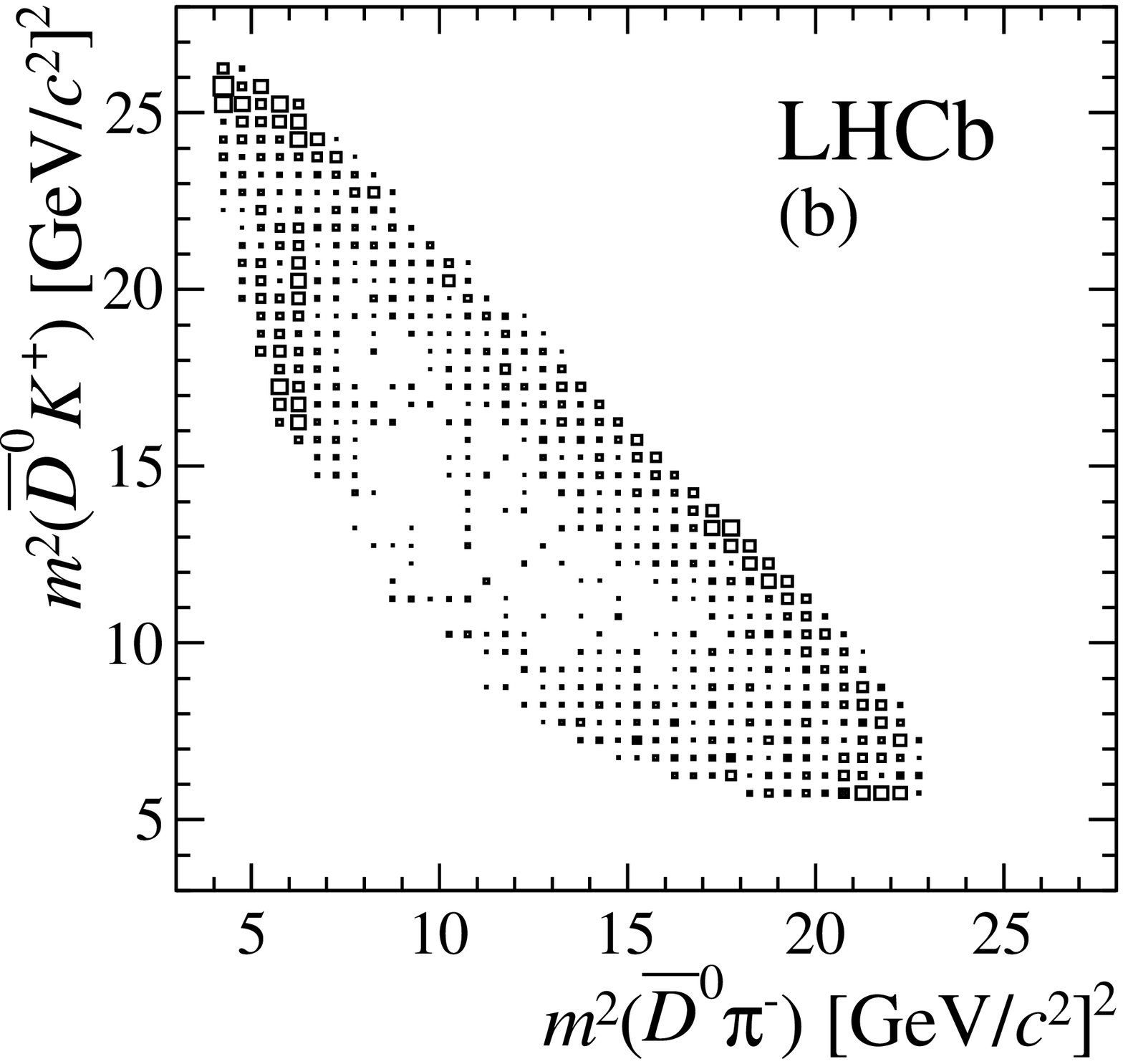}
  \includegraphics[width=0.325\textwidth]{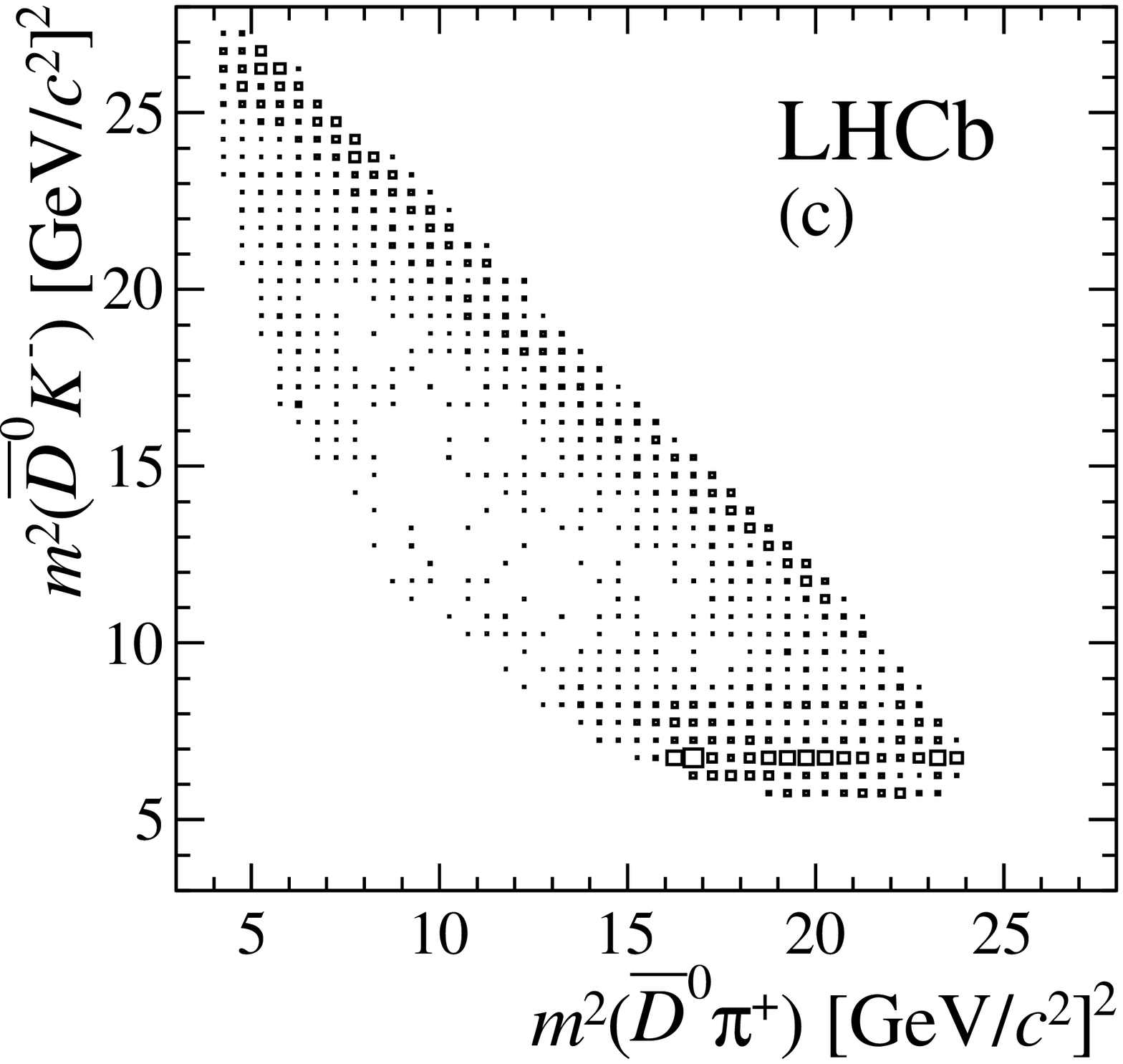}
  \caption{
    \small{Efficiency corrected Dalitz plot distributions for (a)~\mbox{$\Bd\to\Dzb\pip\pim$}, (b)~\mbox{$\Bd\to\Dzb\Kp\pim$} and (c)~\mbox{$\Bs\to\Dzb\Km\pip$} candidates obtained from the signal weights.}
  }
  \label{fig:dalitz}
\end{figure}

The Dalitz plots obtained from the signal weights are shown in Fig.~\ref{fig:dalitz}.
The \mbox{$\Bd\to \Dzb\pip\pim$} plot, Fig.~\ref{fig:dalitz}(a), shows contributions from the $\rho^0(770)$ and $f_2(1270)$ resonances (upper diagonal edge of the Dalitz plot) and from the $\D_2^{*-}(2460)$ state (horizontal band), as expected from previous studies of this decay~\cite{Kuzmin:2006mw,:2010ip}.
The \mbox{$\Bd\to \Dzb\Kp\pim$} plot, Fig.~\ref{fig:dalitz}(b), shows contributions from the $\Kstarz(892)$ (upper diagonal edge) and from the $\D_2^{*-}(2460)$ (vertical band) resonances, also as expected~\cite{Aubert:2005yt}.
The \mbox{$\Bs\to \Dzb\Km\pip$} plot, Fig.~\ref{fig:dalitz}(c), shows contributions from the $\Kstarzb(892)$ (upper diagonal edge) and from the $\D_{s2}^{*-}(2573)$ (horizontal band) states.
The former contribution is as expected~\cite{LHCb-PAPER-2011-008}. The decay \mbox{$\Bs \to \D_{s2}^{*-}(2573) \pip$} has not been observed previously but is expected to exist given the observation of the \mbox{$\Bs \to \D_{s2}^{*-}(2573) \mup \neu X$} decay~\cite{LHCb-PAPER-2011-001}.

\section{Systematic uncertainties and cross-checks}

Systematic uncertainties are assigned to both branching fraction ratios due to the following sources (summarised in Table~\ref{tab:systematics}). 
Note that all uncertainties are relative. 
The variation of efficiency across the Dalitz plot may not be correctly modelled in simulation. 
A two-dimensional polynomial is used to fit the variation across the Dalitz region of each of the four contributions to the efficiency (detector acceptance, selection criteria, PID and trigger effects). 
These polynomials are used to generate 1000 simulated pseudo-experiments, varying the fit parameters within their uncertainties. Each set of simulations is used to calculate the 
efficiency corrected yield. The standard deviation from a Gaussian fit to these yields 
is used to provide a systematic uncertainty for each decay mode. This leads to a systematic uncertainty of 3.4\,\% (3.1\,\%) for $R_{\Bd}$ ($R_{\Bs}$).
The $DK\pi$ fit model is varied by scaling the signal PDF width ratio to account for the different masses of the \Bd and \Bs mesons, replacing the PDFs of the background components with unsmoothed versions, adding components for potential background from \mbox{$\Bs \to \Dstarzb\Kstarzb$} and \mbox{$\Lbbar \to \Dzb \antiproton \pip$} decays, and replacing the double Gaussian signal components with double Crystal Ball~\cite{Skwarnicki:1986xj} functions.  
The $D\pi\pi$ fit model is varied by replacing the PDF of the \mbox{$\Lbbar \to \Dzb \antiproton \pip$} component with an unsmoothed version, varying the slope of the combinatorial background and replacing the exponential partially reconstructed background component with a PDF for \mbox{$\Bd \to \Dstarzb\pip\pim$} decays.
Combined in quadrature, these contribute 6.3\,\% (4.3\,\%) to $R_{\Bd}$ ($R_{\Bs}$).
Variations in the $\Dstarpm$, $\Dpm$ and $\Dspm$ vetoes contribute to $R_{\Bd}$ ($R_{\Bs}$), at the level of $<$0.1\,\%, 2.0\,\% and 0.2\,\% (1.0\,\%, 0.5\,\% and 0.2\,\%), respectively. 
In addition, the possible differences in the data to simulation ratios of trigger and PID efficiencies between the two channels (both 1.0\,\%) and the limited statistics of the simulated data samples used to 
calculate efficiencies (2.0\,\%) affect both $R_{\Bd}$ and $R_{\Bs}$. 
The uncertainty on the quantity $f_s/f_d$ (7.8\,\%) affects only $R_{\Bs}$.
The total systematic uncertainties are obtained as the quadratic sums of all contributions.

\begin{table}[!tb]
  \caption{
    \small{Systematic uncertainties on $R_{\Bd}$ and $R_{\Bs}$.
    The total is obtained from the sum in quadrature of all contributions.
    Note that all uncertainties are relative.}
  }
  \label{tab:systematics}
  \centering
  \begin{tabular}{lcc}
    & \multicolumn{2}{c}{Uncertainty (\%)} \\
    Source & \phantom{$<$} $\Bd$ & $\Bs$ \\ 
    \hline
    Modelling of efficiency & \phantom{$<$} 3.4 & 3.1\\
    Fit model & \phantom{$<$} 6.3 & 4.3\\
    $\Dstarpm$ veto & $<$ 0.1 & 1.0\\
    $\Dpm$ veto & \phantom{$<$} 2.0 & 0.2\\
    $\Dspm$ veto & \phantom{$<$} 0.2 & 0.5\\
    Trigger & \phantom{$<$} 1.0 & 1.0\\
    Particle identification & \phantom{$<$} 1.0 & 1.0\\
    Simulation statistics & \phantom{$<$} 2.0 & 2.0\\
    $f_s/f_d$ & \phantom{$<$} -- & 7.8 \\
    \hline
    Total & \phantom{$<$} 7.8 & 9.8\\
  \end{tabular}
\end{table}

A number of cross-checks are performed to test the stability of the results. 
Based upon the hardware trigger decision, candidates are separated into three groups: 
events in which a particle from the signal decay created a cluster with enough energy in the calorimeter to fire the trigger, events that were triggered independently of the signal decay and those events that were triggered by both the signal decay and the rest of the event. 
The data sample is divided by dipole magnet polarity. 
The neural network and PID requirements are both tightened and loosened. 
The PID efficiency is evaluated using the kinematic properties from $\Dzb\pip\pim$ data instead of from simulation. 
The requirement for the $\Bds$ mass constrained fits to converge is removed. 
All cross-checks give consistent results.

\section{Results and conclusions}

In summary, the decay \mbox{$\Bs\to \Dzb\Km\pip$} has been observed for the first time, and its branching fraction relative to that of the \mbox{$\Bd\to\Dzb\pip\pim$} decay is measured to be
\begin{equation}
  \frac{
    {\cal B}\left(\Bs \to \Dzb K^-\pi^+\right)}{
    {\cal B}\left(\Bd \to \Dzb \pi^+\pi^-\right)} = 1.18 \pm 0.05\,\text{(stat.)} \pm 0.12\,\text{(syst.)} \, .\nonumber
\end{equation}
The current world average value of \mbox{${\cal B}\left(\Bd\to\Dzb\pip\pim\right) = (8.4 \pm 0.4 \pm 0.8) \times 10^{-4}$}~\cite{Kuzmin:2006mw} assumes equal production of 
$\Bp\Bm$ and $\Bz\Bzb$ at the $\FourS$ resonance and uses the $\Dz$ branching fraction \mbox{${\cal B}\left(D^0 \to K^-\pi^+\right) = (3.80 \pm 0.07)\,\%$}. 
Using the current world average values of \mbox{$\Gamma(\FourS\to\Bp\Bm)/\Gamma(\FourS\to\Bz\Bzb) = 1.055 \pm 0.025$}~\cite{PDG2012} 
and \mbox{${\cal B}\left(D^0 \to K^-\pi^+\right) = (3.88 \pm 0.05)\,\%$}~\cite{PDG2012}, 
the branching fraction of the normalisation channel becomes \mbox{${\cal B}\left(\Bd\to\Dzb\pip\pim\right) = (8.5 \pm 0.4 \pm 0.8) \times 10^{-4}$}. This corrected value gives
\begin{equation}
  {\cal B}\left(\Bs\to \Dzb K^-\pi^+\right) = (1.00 \pm 0.04\,\text{(stat.)} \pm 0.10\,\text{(syst.)} \pm 0.10\,\text{(}{\cal B}\text{)})\times 10^{-3} \,,\nonumber
\end{equation}
where the third uncertainty arises from ${\cal B}\left(B^0 \to \Dzb \pi^+\pi^-\right)$.
The \mbox{$\Bd \to \Dzb\Kp\pim$} decay has also been measured,
with relative branching fraction
\begin{equation}
\frac{
  {\cal B}\left(\Bd \to \Dzb K^+\pi^-\right)}{
  {\cal B}\left(\Bd \to \Dzb \pip\pim\right)} = 0.106 \pm 0.007\,\text{(stat.)} \pm 0.008\,\text{(syst.)} \, .\nonumber
\end{equation}
Using the corrected value of ${\cal B}\left(\Bd\to\Dzb\pip\pim\right)$ gives 
\begin{equation}
  {\cal B}\left(\Bd\to \Dzb K^+\pi^-\right) = (9.0 \pm 0.6\,\text{(stat.)} \pm 0.7\,\text{(syst.)} \pm 0.9\,\text{(}{\cal B}\text{)})\times 10^{-5} \,,\nonumber
\end{equation}
which is the most precise measurement of this quantity to date.
Future studies of the Dalitz plot distributions of these decays will provide insight into the dynamics of hadronic $\B$ decays.  
In addition, the \mbox{$\Bd \to \D\Kp\pim$} decay may be used to measure the $\CP$ violating phase $\gamma$.

\section*{Acknowledgements}

\noindent We express our gratitude to our colleagues in the CERN
accelerator departments for the excellent performance of the LHC. We
thank the technical and administrative staff at the LHCb
institutes. We acknowledge support from CERN and from the national
agencies: CAPES, CNPq, FAPERJ and FINEP (Brazil); NSFC (China);
CNRS/IN2P3 and Region Auvergne (France); BMBF, DFG, HGF and MPG
(Germany); SFI (Ireland); INFN (Italy); FOM and NWO (The Netherlands);
SCSR (Poland); ANCS/IFA (Romania); MinES, Rosatom, RFBR and NRC
``Kurchatov Institute'' (Russia); MinECo, XuntaGal and GENCAT (Spain);
SNSF and SER (Switzerland); NAS Ukraine (Ukraine); STFC (United
Kingdom); NSF (USA). We also acknowledge the support received from the
ERC under FP7. The Tier1 computing centres are supported by IN2P3
(France), KIT and BMBF (Germany), INFN (Italy), NWO and SURF (The
Netherlands), PIC (Spain), GridPP (United Kingdom). We are thankful
for the computing resources put at our disposal by Yandex LLC
(Russia), as well as to the communities behind the multiple open
source software packages that we depend on.

\ifx\mcitethebibliography\mciteundefinedmacro
\PackageError{LHCb.bst}{mciteplus.sty has not been loaded}
{This bibstyle requires the use of the mciteplus package.}\fi
\providecommand{\href}[2]{#2}


\begin{mcitethebibliography}{10}
\mciteSetBstSublistMode{n}
\mciteSetBstMaxWidthForm{subitem}{\alph{mcitesubitemcount})}
\mciteSetBstSublistLabelBeginEnd{\mcitemaxwidthsubitemform\space}
{\relax}{\relax}

\bibitem{Cabibbo:1963yz}
N.~Cabibbo, \ifthenelse{\boolean{articletitles}}{{\it {Unitary symmetry and
  leptonic decays}},
  }{}\href{http://dx.doi.org/10.1103/PhysRevLett.10.531}{Phys.\ Rev.\ Lett.\
  {\bf 10} (1963) 531}\relax
\mciteBstWouldAddEndPuncttrue
\mciteSetBstMidEndSepPunct{\mcitedefaultmidpunct}
{\mcitedefaultendpunct}{\mcitedefaultseppunct}\relax
\EndOfBibitem
\bibitem{Kobayashi:1973fv}
M.~Kobayashi and T.~Maskawa, \ifthenelse{\boolean{articletitles}}{{\it {CP
  violation in the renormalizable theory of weak interaction}},
  }{}\href{http://dx.doi.org/10.1143/PTP.49.652}{Prog.\ Theor.\ Phys.\  {\bf
  49} (1973) 652}\relax
\mciteBstWouldAddEndPuncttrue
\mciteSetBstMidEndSepPunct{\mcitedefaultmidpunct}
{\mcitedefaultendpunct}{\mcitedefaultseppunct}\relax
\EndOfBibitem
\bibitem{:2013zd}
BaBar collaboration, J.~P. Lees {\em et~al.},
  \ifthenelse{\boolean{articletitles}}{{\it {Observation of direct \CP
  violation in the measurement of the Cabibbo-Kobayashi-Maskawa angle $\gamma$
  with $\Bpm \to D^{(*)}K^{(*)\pm}$ decays}},
  }{}\href{http://dx.doi.org/10.1103/PhysRevD.87.052015}{Phys.\ Rev.\  {\bf
  D87} (2013) 052015}, \href{http://arxiv.org/abs/1301.1029}{{\tt
  arXiv:1301.1029}}\relax
\mciteBstWouldAddEndPuncttrue
\mciteSetBstMidEndSepPunct{\mcitedefaultmidpunct}
{\mcitedefaultendpunct}{\mcitedefaultseppunct}\relax
\EndOfBibitem
\bibitem{Trabelsi:2013uj}
K.~Trabelsi, \ifthenelse{\boolean{articletitles}}{{\it {Study of direct \CP in
  charmed \B decays and measurement of the CKM angle $\gamma$ at Belle}},
  }{}\href{http://arxiv.org/abs/1301.2033}{{\tt arXiv:1301.2033}}, {proceedings
  of CKM 2012, the $7^{\rm th}$ International Workshop on the CKM Unitarity
  Triangle}\relax
\mciteBstWouldAddEndPuncttrue
\mciteSetBstMidEndSepPunct{\mcitedefaultmidpunct}
{\mcitedefaultendpunct}{\mcitedefaultseppunct}\relax
\EndOfBibitem
\bibitem{LHCb-PAPER-2013-020}
LHCb collaboration, R.~Aaij {\em et~al.},
  \ifthenelse{\boolean{articletitles}}{{\it {A measurement of $\gamma$ from a
  combination of $\Bpm \to Dh^\pm$ analyses}},
  }{}\href{http://arxiv.org/abs/1305.2050}{{\tt arXiv:1305.2050}}, {submitted
  to Phys. Lett. B}\relax
\mciteBstWouldAddEndPuncttrue
\mciteSetBstMidEndSepPunct{\mcitedefaultmidpunct}
{\mcitedefaultendpunct}{\mcitedefaultseppunct}\relax
\EndOfBibitem
\bibitem{Gronau:2002mu}
M.~Gronau, \ifthenelse{\boolean{articletitles}}{{\it {Improving bounds on
  $\gamma$ in $B^\pm \to DK^\pm$ and $B^{\pm,0} \to DX_{s}^{\pm,0}$}},
  }{}\href{http://dx.doi.org/10.1016/S0370-2693(03)00192-8}{Phys.\ Lett.\  {\bf
  B557} (2003) 198}, \href{http://arxiv.org/abs/hep-ph/0211282}{{\tt
  arXiv:hep-ph/0211282}}\relax
\mciteBstWouldAddEndPuncttrue
\mciteSetBstMidEndSepPunct{\mcitedefaultmidpunct}
{\mcitedefaultendpunct}{\mcitedefaultseppunct}\relax
\EndOfBibitem
\bibitem{LHCb-PAPER-2012-042}
LHCb collaboration, R.~Aaij {\em et~al.},
  \ifthenelse{\boolean{articletitles}}{{\it {Measurement of \CP observables in
  $B^0\to D K^{*0}$ with $D\to K^+K^-$}},
  }{}\href{http://dx.doi.org/10.1007/JHEP03(2013)067}{JHEP {\bf 03} (2013) 67},
  \href{http://arxiv.org/abs/1212.5205}{{\tt arXiv:1212.5205}}\relax
\mciteBstWouldAddEndPuncttrue
\mciteSetBstMidEndSepPunct{\mcitedefaultmidpunct}
{\mcitedefaultendpunct}{\mcitedefaultseppunct}\relax
\EndOfBibitem
\bibitem{Gershon:2008pe}
T.~Gershon, \ifthenelse{\boolean{articletitles}}{{\it {On the measurement of
  the Unitarity Triangle angle $\gamma$ from $B^0 \to DK^{*0}$ decays}},
  }{}\href{http://dx.doi.org/10.1103/PhysRevD.79.051301}{Phys.\ Rev.\  {\bf
  D79} (2009) 051301}, \href{http://arxiv.org/abs/0810.2706}{{\tt
  arXiv:0810.2706}}\relax
\mciteBstWouldAddEndPuncttrue
\mciteSetBstMidEndSepPunct{\mcitedefaultmidpunct}
{\mcitedefaultendpunct}{\mcitedefaultseppunct}\relax
\EndOfBibitem
\bibitem{Gershon:2009qc}
T.~Gershon and M.~Williams, \ifthenelse{\boolean{articletitles}}{{\it
  {Prospects for the measurement of the Unitarity Triangle angle $\gamma$ from
  $B^0 \to DK^+ \pi^-$ decays}},
  }{}\href{http://dx.doi.org/10.1103/PhysRevD.80.092002}{Phys.\ Rev.\  {\bf
  D80} (2009) 092002}, \href{http://arxiv.org/abs/0909.1495}{{\tt
  arXiv:0909.1495}}\relax
\mciteBstWouldAddEndPuncttrue
\mciteSetBstMidEndSepPunct{\mcitedefaultmidpunct}
{\mcitedefaultendpunct}{\mcitedefaultseppunct}\relax
\EndOfBibitem
\bibitem{Aubert:2005yt}
BaBar collaboration, B.~Aubert {\em et~al.},
  \ifthenelse{\boolean{articletitles}}{{\it {Measurement of branching fractions
  and resonance contributions for $B^0 \to \bar{D}^0 K^{+} \pi^{-}$ and search
  for $B^0 \to D^0 K^{+} \pi^{-}$ decays}},
  }{}\href{http://dx.doi.org/10.1103/PhysRevLett.96.011803}{Phys.\ Rev.\ Lett.\
   {\bf 96} (2006) 011803}, \href{http://arxiv.org/abs/hep-ex/0509036}{{\tt
  arXiv:hep-ex/0509036}}\relax
\mciteBstWouldAddEndPuncttrue
\mciteSetBstMidEndSepPunct{\mcitedefaultmidpunct}
{\mcitedefaultendpunct}{\mcitedefaultseppunct}\relax
\EndOfBibitem
\bibitem{LHCb-PAPER-2011-008}
LHCb collaboration, R.~Aaij {\em et~al.},
  \ifthenelse{\boolean{articletitles}}{{\it {First observation of the decay
  $\overline{B}^0_s \to D^0 K^{*0}$ and a measurement of the ratio of branching
  fractions $\frac{{\cal B}(\overline{B}^0_s \to D^0 K^{*0})}{{\cal
  B}(\overline{B}^0 \to D^0 \rho^0)}$}},
  }{}\href{http://dx.doi.org/10.1016/j.physletb.2011.10.073}{Phys.\ Lett.\
  {\bf B706} (2011) 32}, \href{http://arxiv.org/abs/1110.3676}{{\tt
  arXiv:1110.3676}}\relax
\mciteBstWouldAddEndPuncttrue
\mciteSetBstMidEndSepPunct{\mcitedefaultmidpunct}
{\mcitedefaultendpunct}{\mcitedefaultseppunct}\relax
\EndOfBibitem
\bibitem{Kuzmin:2006mw}
Belle collaboration, A.~Kuzmin {\em et~al.},
  \ifthenelse{\boolean{articletitles}}{{\it {Study of $\bar{B}^0 \to D^0 \pi^+
  \pi^-$ decays}},
  }{}\href{http://dx.doi.org/10.1103/PhysRevD.76.012006}{Phys.\ Rev.\  {\bf
  D76} (2007) 012006}, \href{http://arxiv.org/abs/hep-ex/0611054}{{\tt
  arXiv:hep-ex/0611054}}\relax
\mciteBstWouldAddEndPuncttrue
\mciteSetBstMidEndSepPunct{\mcitedefaultmidpunct}
{\mcitedefaultendpunct}{\mcitedefaultseppunct}\relax
\EndOfBibitem
\bibitem{:2010ip}
BaBar collaboration, P.~del Amo~Sanchez {\em et~al.},
  \ifthenelse{\boolean{articletitles}}{{\it {Dalitz-plot analysis of $B^0 \to
  \bar{D}^0 \pi^+ \pi^-$}}, }{}PoS {\bf ICHEP2010} (2010) 250,
  \href{http://arxiv.org/abs/1007.4464}{{\tt arXiv:1007.4464}}\relax
\mciteBstWouldAddEndPuncttrue
\mciteSetBstMidEndSepPunct{\mcitedefaultmidpunct}
{\mcitedefaultendpunct}{\mcitedefaultseppunct}\relax
\EndOfBibitem
\bibitem{Alves:2008zz}
LHCb collaboration, A.~A. Alves~Jr. {\em et~al.},
  \ifthenelse{\boolean{articletitles}}{{\it {The \lhcb detector at the LHC}},
  }{}\href{http://dx.doi.org/10.1088/1748-0221/3/08/S08005}{JINST {\bf 3}
  (2008) S08005}\relax
\mciteBstWouldAddEndPuncttrue
\mciteSetBstMidEndSepPunct{\mcitedefaultmidpunct}
{\mcitedefaultendpunct}{\mcitedefaultseppunct}\relax
\EndOfBibitem
\bibitem{LHCb-DP-2012-003}
M.~Adinolfi {\em et~al.}, \ifthenelse{\boolean{articletitles}}{{\it
  {Performance of the \lhcb RICH detector at the LHC}},
  }{}\href{http://dx.doi.org/10.1140/epjc/s10052-013-2431-9}{Eur.\ Phys.\ J.\
  {\bf C73} (2013) 2431}, \href{http://arxiv.org/abs/1211.6759}{{\tt
  arXiv:1211.6759}}\relax
\mciteBstWouldAddEndPuncttrue
\mciteSetBstMidEndSepPunct{\mcitedefaultmidpunct}
{\mcitedefaultendpunct}{\mcitedefaultseppunct}\relax
\EndOfBibitem
\bibitem{LHCb-DP-2012-004}
R.~Aaij {\em et~al.}, \ifthenelse{\boolean{articletitles}}{{\it {The \lhcb
  trigger and its performance in 2011}},
  }{}\href{http://dx.doi.org/10.1088/1748-0221/8/04/P04022}{JINST {\bf 8}
  (2013) P04022}, \href{http://arxiv.org/abs/1211.3055}{{\tt
  arXiv:1211.3055}}\relax
\mciteBstWouldAddEndPuncttrue
\mciteSetBstMidEndSepPunct{\mcitedefaultmidpunct}
{\mcitedefaultendpunct}{\mcitedefaultseppunct}\relax
\EndOfBibitem
\bibitem{Gligorov:2012qt}
V.~V. Gligorov and M.~Williams, \ifthenelse{\boolean{articletitles}}{{\it
  {Efficient, reliable and fast high-level triggering using a bonsai boosted
  decision tree}},
  }{}\href{http://dx.doi.org/10.1088/1748-0221/8/02/P02013}{JINST {\bf 8}
  (2013) P02013}, \href{http://arxiv.org/abs/1210.6861}{{\tt
  arXiv:1210.6861}}\relax
\mciteBstWouldAddEndPuncttrue
\mciteSetBstMidEndSepPunct{\mcitedefaultmidpunct}
{\mcitedefaultendpunct}{\mcitedefaultseppunct}\relax
\EndOfBibitem
\bibitem{LHCb-PAPER-2012-018}
LHCb collaboration, R.~Aaij {\em et~al.},
  \ifthenelse{\boolean{articletitles}}{{\it {Observation of $B^0 \to
  \overline{D}^0 K^+ K^-$ and evidence for $B^0_s \to \overline{D}^0 K^+
  K^-$}}, }{}\href{http://dx.doi.org/10.1103/PhysRevLett.109.131801}{Phys.\
  Rev.\ Lett.\  {\bf 109} (2012) 131801},
  \href{http://arxiv.org/abs/1207.5991}{{\tt arXiv:1207.5991}}\relax
\mciteBstWouldAddEndPuncttrue
\mciteSetBstMidEndSepPunct{\mcitedefaultmidpunct}
{\mcitedefaultendpunct}{\mcitedefaultseppunct}\relax
\EndOfBibitem
\bibitem{Sjostrand:2006za}
T.~Sj\"{o}strand, S.~Mrenna, and P.~Skands,
  \ifthenelse{\boolean{articletitles}}{{\it {PYTHIA 6.4 physics and manual}},
  }{}\href{http://dx.doi.org/10.1088/1126-6708/2006/05/026}{JHEP {\bf 05}
  (2006) 026}, \href{http://arxiv.org/abs/hep-ph/0603175}{{\tt
  arXiv:hep-ph/0603175}}\relax
\mciteBstWouldAddEndPuncttrue
\mciteSetBstMidEndSepPunct{\mcitedefaultmidpunct}
{\mcitedefaultendpunct}{\mcitedefaultseppunct}\relax
\EndOfBibitem
\bibitem{LHCb-PROC-2010-056}
I.~Belyaev {\em et~al.}, \ifthenelse{\boolean{articletitles}}{{\it {Handling of
  the generation of primary events in \gauss, the \lhcb simulation framework}},
  }{}\href{http://dx.doi.org/10.1109/NSSMIC.2010.5873949}{Nuclear Science
  Symposium Conference Record (NSS/MIC) {\bf IEEE} (2010) 1155}\relax
\mciteBstWouldAddEndPuncttrue
\mciteSetBstMidEndSepPunct{\mcitedefaultmidpunct}
{\mcitedefaultendpunct}{\mcitedefaultseppunct}\relax
\EndOfBibitem
\bibitem{Lange:2001uf}
D.~J. Lange, \ifthenelse{\boolean{articletitles}}{{\it {The EvtGen particle
  decay simulation package}},
  }{}\href{http://dx.doi.org/10.1016/S0168-9002(01)00089-4}{Nucl.\ Instrum.\
  Meth.\  {\bf A462} (2001) 152}\relax
\mciteBstWouldAddEndPuncttrue
\mciteSetBstMidEndSepPunct{\mcitedefaultmidpunct}
{\mcitedefaultendpunct}{\mcitedefaultseppunct}\relax
\EndOfBibitem
\bibitem{Golonka:2005pn}
P.~Golonka and Z.~Was, \ifthenelse{\boolean{articletitles}}{{\it {PHOTOS Monte
  Carlo: a precision tool for QED corrections in $Z$ and $W$ decays}},
  }{}\href{http://dx.doi.org/10.1140/epjc/s2005-02396-4}{Eur.\ Phys.\ J.\  {\bf
  C45} (2006) 97}, \href{http://arxiv.org/abs/hep-ph/0506026}{{\tt
  arXiv:hep-ph/0506026}}\relax
\mciteBstWouldAddEndPuncttrue
\mciteSetBstMidEndSepPunct{\mcitedefaultmidpunct}
{\mcitedefaultendpunct}{\mcitedefaultseppunct}\relax
\EndOfBibitem
\bibitem{Allison:2006ve}
GEANT4 collaboration, J.~Allison {\em et~al.},
  \ifthenelse{\boolean{articletitles}}{{\it {Geant4 developments and
  applications}}, }{}\href{http://dx.doi.org/10.1109/TNS.2006.869826}{IEEE
  Trans.\ Nucl.\ Sci.\  {\bf 53} (2006) 270}\relax
\mciteBstWouldAddEndPuncttrue
\mciteSetBstMidEndSepPunct{\mcitedefaultmidpunct}
{\mcitedefaultendpunct}{\mcitedefaultseppunct}\relax
\EndOfBibitem
\bibitem{Agostinelli:2002hh}
GEANT4 collaboration, S.~Agostinelli {\em et~al.},
  \ifthenelse{\boolean{articletitles}}{{\it {GEANT4: A simulation toolkit}},
  }{}\href{http://dx.doi.org/10.1016/S0168-9002(03)01368-8}{Nucl.\ Instrum.\
  Meth.\  {\bf A506} (2003) 250}\relax
\mciteBstWouldAddEndPuncttrue
\mciteSetBstMidEndSepPunct{\mcitedefaultmidpunct}
{\mcitedefaultendpunct}{\mcitedefaultseppunct}\relax
\EndOfBibitem
\bibitem{LHCb-PROC-2011-006}
M.~Clemencic {\em et~al.}, \ifthenelse{\boolean{articletitles}}{{\it {The \lhcb
  simulation application, \gauss: design, evolution and experience}},
  }{}\href{http://dx.doi.org/10.1088/1742-6596/331/3/032023}{{J.\ of Phys.\ :
  Conf.\ Ser.\ } {\bf 331} (2011) 032023}\relax
\mciteBstWouldAddEndPuncttrue
\mciteSetBstMidEndSepPunct{\mcitedefaultmidpunct}
{\mcitedefaultendpunct}{\mcitedefaultseppunct}\relax
\EndOfBibitem
\bibitem{Breiman}
L.~Breiman, J.~H. Friedman, R.~A. Olshen, and C.~J. Stone, {\em Classification
  and regression trees}.
\newblock Wadsworth international group, Belmont, California, USA, 1984\relax
\mciteBstWouldAddEndPuncttrue
\mciteSetBstMidEndSepPunct{\mcitedefaultmidpunct}
{\mcitedefaultendpunct}{\mcitedefaultseppunct}\relax
\EndOfBibitem
\bibitem{LHCb-PAPER-2012-025}
LHCb collaboration, R.~Aaij {\em et~al.},
  \ifthenelse{\boolean{articletitles}}{{\it {First evidence for the
  annihilation decay mode $B^{+} \to D_{s}^{+} \phi$}},
  }{}\href{http://dx.doi.org/10.1007/JHEP02(2013)043}{JHEP {\bf 02} (2013) 43},
  \href{http://arxiv.org/abs/1210.1089}{{\tt arXiv:1210.1089}}\relax
\mciteBstWouldAddEndPuncttrue
\mciteSetBstMidEndSepPunct{\mcitedefaultmidpunct}
{\mcitedefaultendpunct}{\mcitedefaultseppunct}\relax
\EndOfBibitem
\bibitem{LHCb-PAPER-2012-050}
LHCb collaboration, R.~Aaij {\em et~al.},
  \ifthenelse{\boolean{articletitles}}{{\it {First observations of $B^0 \to
  D^+D^-$, $D_s^+D^-$ and $D^0\overline{D}^0$ decays}},
  }{}\href{http://dx.doi.org/10.1103/PhysRevD.87.092007}{Phys.\ Rev.\  {\bf
  D87} (2013) 092007}, \href{http://arxiv.org/abs/1302.5854}{{\tt
  arXiv:1302.5854}}\relax
\mciteBstWouldAddEndPuncttrue
\mciteSetBstMidEndSepPunct{\mcitedefaultmidpunct}
{\mcitedefaultendpunct}{\mcitedefaultseppunct}\relax
\EndOfBibitem
\bibitem{Pivk:2004ty}
M.~Pivk and F.~R. Le~Diberder, \ifthenelse{\boolean{articletitles}}{{\it
  {sPlot: a statistical tool to unfold data distributions}},
  }{}\href{http://dx.doi.org/10.1016/j.nima.2005.08.106}{Nucl.\ Instrum.\
  Meth.\  {\bf A555} (2005) 356},
  \href{http://arxiv.org/abs/physics/0402083}{{\tt
  arXiv:physics/0402083}}\relax
\mciteBstWouldAddEndPuncttrue
\mciteSetBstMidEndSepPunct{\mcitedefaultmidpunct}
{\mcitedefaultendpunct}{\mcitedefaultseppunct}\relax
\EndOfBibitem
\bibitem{Feindt2006190}
M.~Feindt and U.~Kerzel, \ifthenelse{\boolean{articletitles}}{{\it {The
  NeuroBayes neural network package}},
  }{}\href{http://dx.doi.org/10.1016/j.nima.2005.11.166}{Nucl.\ Instrum.\
  Meth.\  {\bf A559} (2006) 190}\relax
\mciteBstWouldAddEndPuncttrue
\mciteSetBstMidEndSepPunct{\mcitedefaultmidpunct}
{\mcitedefaultendpunct}{\mcitedefaultseppunct}\relax
\EndOfBibitem
\bibitem{LHCB-PAPER-2012-001}
LHCb collaboration, R.~Aaij {\em et~al.},
  \ifthenelse{\boolean{articletitles}}{{\it {Observation of \CP violation in
  $\Bpm \to D\Kpm$ decays}},
  }{}\href{http://dx.doi.org/10.1016/j.physletb.2012.04.060}{Phys.\ Lett.\
  {\bf B712} (2012) 203}, \href{http://arxiv.org/abs/1203.3662}{{\tt
  arXiv:1203.3662}}\relax
\mciteBstWouldAddEndPuncttrue
\mciteSetBstMidEndSepPunct{\mcitedefaultmidpunct}
{\mcitedefaultendpunct}{\mcitedefaultseppunct}\relax
\EndOfBibitem
\bibitem{Hulsbergen:2005pu}
W.~D. Hulsbergen, \ifthenelse{\boolean{articletitles}}{{\it {Decay chain
  fitting with a Kalman filter}},
  }{}\href{http://dx.doi.org/10.1016/j.nima.2005.06.078}{Nucl.\ Instrum.\
  Meth.\  {\bf A552} (2005) 566},
  \href{http://arxiv.org/abs/physics/0503191}{{\tt
  arXiv:physics/0503191}}\relax
\mciteBstWouldAddEndPuncttrue
\mciteSetBstMidEndSepPunct{\mcitedefaultmidpunct}
{\mcitedefaultendpunct}{\mcitedefaultseppunct}\relax
\EndOfBibitem
\bibitem{PDG2012}
Particle Data Group, J.~Beringer {\em et~al.},
  \ifthenelse{\boolean{articletitles}}{{\it {\href{http://pdg.lbl.gov/}{Review
  of particle physics}}},
  }{}\href{http://dx.doi.org/10.1103/PhysRevD.86.010001}{Phys.\ Rev.\  {\bf
  D86} (2012) 010001}\relax
\mciteBstWouldAddEndPuncttrue
\mciteSetBstMidEndSepPunct{\mcitedefaultmidpunct}
{\mcitedefaultendpunct}{\mcitedefaultseppunct}\relax
\EndOfBibitem
\bibitem{LHCb-CONF-2011-036}
{LHCb collaboration}, \ifthenelse{\boolean{articletitles}}{{\it {Studies of
  beauty baryons decaying to $D^{0}p\pi^{-}$ and $D^{0}pK^{-}$}}, }{}
  \href{http://cdsweb.cern.ch/search?p={LHCb-CONF-2011-036}&f=reportnumber&act%
ion_search=Search&c=LHCb+Reports&c=LHCb+Conference+Proceedings&c=LHCb+Conferen%
ce+Contributions&c=LHCb+Notes&c=LHCb+Theses&c=LHCb+Papers}
  {{LHCb-CONF-2011-036}}\relax
\mciteBstWouldAddEndPuncttrue
\mciteSetBstMidEndSepPunct{\mcitedefaultmidpunct}
{\mcitedefaultendpunct}{\mcitedefaultseppunct}\relax
\EndOfBibitem
\bibitem{LHCb-PAPER-2012-037}
LHCb collaboration, R.~Aaij {\em et~al.},
  \ifthenelse{\boolean{articletitles}}{{\it {Measurement of the fragmentation
  fraction ratio $f_s/f_d$ and its dependence on $B$ meson kinematics}},
  }{}\href{http://dx.doi.org/10.1007/JHEP04(2013)001}{JHEP {\bf 04} (2013) 1},
  \href{http://arxiv.org/abs/1301.5286}{{\tt arXiv:1301.5286}}\relax
\mciteBstWouldAddEndPuncttrue
\mciteSetBstMidEndSepPunct{\mcitedefaultmidpunct}
{\mcitedefaultendpunct}{\mcitedefaultseppunct}\relax
\EndOfBibitem
\bibitem{LHCb-PAPER-2012-002}
LHCb collaboration, R.~Aaij {\em et~al.},
  \ifthenelse{\boolean{articletitles}}{{\it {Measurement of $b$-hadron
  branching fractions for two-body decays into charmless charged hadrons}},
  }{}\href{http://dx.doi.org/10.1007/JHEP10(2012)037}{JHEP {\bf 10} (2012) 37},
  \href{http://arxiv.org/abs/1206.2794}{{\tt arXiv:1206.2794}}\relax
\mciteBstWouldAddEndPuncttrue
\mciteSetBstMidEndSepPunct{\mcitedefaultmidpunct}
{\mcitedefaultendpunct}{\mcitedefaultseppunct}\relax
\EndOfBibitem
\bibitem{LHCb-PAPER-2011-001}
LHCb collaboration, R.~Aaij {\em et~al.},
  \ifthenelse{\boolean{articletitles}}{{\it {First observation of $\Bsb \to
  D_{s2}^{*+} X \mun\neub$ decays}},
  }{}\href{http://dx.doi.org/10.1016/j.physletb.2011.02.039}{Phys.\ Lett.\
  {\bf B698} (2011) 14}, \href{http://arxiv.org/abs/1102.0348}{{\tt
  arXiv:1102.0348}}\relax
\mciteBstWouldAddEndPuncttrue
\mciteSetBstMidEndSepPunct{\mcitedefaultmidpunct}
{\mcitedefaultendpunct}{\mcitedefaultseppunct}\relax
\EndOfBibitem
\bibitem{Skwarnicki:1986xj}
T.~Skwarnicki, {\em {A study of the radiative cascade transitions between the
  Upsilon-prime and Upsilon resonances}}, PhD thesis, Institute of Nuclear
  Physics, Krakow, 1986,
  {\href{http://inspirehep.net/record/230779/files/230779.pdf}{DESY-F31-86-02}%
}\relax
\mciteBstWouldAddEndPuncttrue
\mciteSetBstMidEndSepPunct{\mcitedefaultmidpunct}
{\mcitedefaultendpunct}{\mcitedefaultseppunct}\relax
\EndOfBibitem
\end{mcitethebibliography}
\end{document}